\documentclass[aps,pra,preprint,showpacs,amsmath,amssymb]{revtex4}
\input{epsf}

\usepackage{graphicx}
\usepackage{longtable}
\begin{document}

\title{\bf
The dynamic dipole polarizabilities of the Li atom and the Be$^{+}$ ion}
\author{Li-Yan Tang$^{1,2}$, Zong-Chao Yan$^{3,4}$, Ting-Yun Shi$^{1}$, and J. Mitroy$^{5}$}

\affiliation {$^1$State Key Laboratory of Magnetic Resonance and
Atomic and Molecular Physics, Wuhan Institute of Physics and
Mathematics, Chinese Academy of Sciences, Wuhan 430071, P. R. China}

\affiliation {$^{2}$Graduate School of the Chinese Academy of
Sciences, Beijing 100049, P. R. China }

\affiliation {$^3$ Center for Cold Atom Physics, Chinese Academy of
Sciences, Wuhan 430071, P. R. China}

\affiliation{$^4$Department of Physics, Wuhan University, Wuhan
430072, China  and Department of Physics, University of New
Brunswick, Fredericton, New Brunswick, Canada E3B 5A3}

\affiliation {$^{5}$School of Engineering, Charles Darwin
University, Darwin NT 0909, Australia}
\date{\today}

\begin{abstract}
{The dynamic dipole polarizabilities for the Li atom and the Be$^+$
ion in the $2\,^2\!S$ and $2\,^2\!P$ states are calculated using the
variational method with a Hylleraas basis. The present
polarizabilities represent the definitive values in the
non-relativistic limit. Corrections due to relativistic effects are
also estimated. Analytic representations of the polarizabilities for
frequency ranges encompassing the $n=3$ excitations are presented.
The recommended polarizabilities for $^7$Li and $^9$Be$^+$ were
$164.11\pm 0.03$ $a_0^3$ and $24.489 \pm 0.004$ $a_0^3$. }
\end{abstract}

\pacs{31.15.ac, 31.15.ap, 34.20.Cf} \maketitle

\section{Introduction}

The advent of cold atom physics has lead to increased importance
being given to the precise determination atomic polarizabilities and
related quantities. One very important source of systematic error in
the new generation of atomic frequency standards is the blackbody
radiation (BBR) shift~\cite{Rosenband,Margolis,Chwalla}. The
differential Stark shifts caused by the ambient electromagnetic
field leads to a temperature dependent shift in the transition
frequency of the two states involved in the clock transition. The
dynamic polarizability is also useful in the determination of the
magic wavelength in optical lattices
\cite{barber,katori,stan,inouye}. Another area where polarization
phenomena is important is in the determination of global potential
surfaces for diatomic molecules~\cite{leroy}.

When consideration is given to all the atoms and ions commonly used
in cold atom physics, the Li atom and Be$^+$ ion have the advantage
that they have only three electrons. This makes them accessible to
calculations using correlated basis sets with the consequence that
many properties of these systems can be computed to a high degree of
precision. The results of these first principle calculations can
serve as atomic based standards for quantities that are not amenable
to precision measurement. For example, cold-atom interferometry has
been used to measure the ground state polarizabilities of the Li and
Na atoms \cite{miffre,ekstrom}. However the polarizability ratio,
$\alpha_d(X)/\alpha_d(\text{Li})$ can be measured to a higher degree
of precision than the individual polarizabilities~\cite{cronin}. So
measurements of this ratio, in conjunction with a high precision
{\em ab-initio} calculation could lead to a new level of accuracy in
polarizability measurements for the atomic species most commonly
used in cold-atom physics.

Calculations and measurements of Stark shifts are particularly
important in atomic clock research since the BBR shift is
predominantly determined by the Stark shift of the two levels
involved in the clock transition. The best experimental measurements
of the Stark shift have been carried out for the alkali atoms and
accuracies better than 0.1$\%$ have been reported
~\cite{miller,hunter}. Experimental work at this level of accuracy
relies on a very precise determination of the electric field
strength in the interaction
region~\cite{hunter,hunter92,wijngaarden}. High precision Hylleraas
calculations of the type presented here provide an invaluable test
of the experimental reliability since they provide an independent
means for the calibration of electric fields~\cite{stevens}.

The dynamic Stark shift in oscillating electromagnetic fields is
also of interest. The so-called magic wavelength, i.e. the precise
wavelength at which the Stark shifts for upper and lower levels of
the clock transition are the same, is an important parameter for
optical lattices. The present calculation is used to estimate the
magic wavelength for the Li $2\,^2\!S$ $\to$ $2\,^2\!P$ transition.
The present calculations of the AC Stark shift potentially provides
an atomic based standard of electromagnetic (EM) field intensity for
finite frequency radiation.

There have been many calculations of the static polarizabilities of
the ground and excited states of the Li atom and the Be$^+$ ion
\cite{tang,yan,yan1,cohen,zhang,johnson,tang2}. The most precise
calculations on Li and Be$^+$ are the Hylleraas calculations by Tang
and collaborators \cite{tang,tang2}. The Hylleraas calculations were
non-relativistic and also included finite mass effects for Li. Large
scale calculations using fully correlated Hylleraas basis sets can
attain a degree of precision not possible for calculations based on
orbital basis sets \cite{McKDra91,yan1,yan-drake97}. There have been
many calculations of the dynamic polarizability for Li
\cite{merawa94,merawa98,pipin,cohen,chernov,kobayashi,muszynska,safronova},
but fewer for Be$^+$ \cite{merawa98,muszynska}. The present
calculation is by far the most precise calculation of the dynamic
polarizability that is based upon a solution of the non-relativistic
Schr\"{o}dinger equation. One particularly noteworthy treatment is
the relativistic single-double all-order many body perturbation
theory calculation (MBPT-SD) by Safronova \textit{et al.}
\cite{safronova}. This calculation is fully relativistic and treats
correlation effects to a high level of accuracy, although it does
not achieve the same level of precision as the present Hylleraas
calculation.

The present work computes the dynamic dipole polarizabilities of the
Li atom and the Be$^+$ ion in the $2\,^2\!S$, and $2\,^2\!P$ levels
using a large variational calculation with a Hylleraas basis set.
This methodology allows for the determination of the computational
uncertainty related to the convergence of the basis set. Analytic
representations of the dynamic polarizabilities are made so they can
subsequently be computed at any frequency. Finally, the difference
between the calculated and experimental binding energies is used to
estimate the size of the relativistic correction to the
polarizability. The final polarizabilities should be regarded as the
recommended polarizabilities for comparison with experiment. All
quantities given in this work are reported in atomic units except
where indicated otherwise.

\begin{table}
\caption{Comparisons of the binding energies (in a.u.) of Li and
Be$^+$ in their low-lying states. The experimental valence binding
energies are taken from the National Institute of Standards
database~\cite{ralchenko}. The $J$-weighted average is used for
states with $L \ge 1$.  The ground-state energies for the
$^{\infty}$Li$^+$ and $^{\infty}$Be$^{2+}$ ions are
$-7.2799134126693059$ and $-13.6555662384235867$ a.u. respectively
~\cite{drake}. The ground-state energies for $^7$Li$^+$ and
$^9$Be$^{2+}$ are $-7.2793215198156744$ and $-13.6547092682827917$
a.u. respectively~\cite{drake}. Underlining is used to indicate
digits that have not converged with respect to basis set
enlargement.} \label{tab:1}
\begin{ruledtabular}
\begin{tabular}{lccc}
\multicolumn{1}{l}{State} & \multicolumn{2}{c}{Theory} & \multicolumn{1}{c}{Experiment} \\
  \cline{2-3}
    & \multicolumn{1}{c}{$^{\infty}$Li} & \multicolumn{1}{c}{$^{7}$Li}  & \multicolumn{1}{c}{$^{6,7}$Li} \\
$2\,^2\!S$ &$-0.19814691\underline{124}$  &$-0.19813041\underline{084}$ &--0.198142 \\
$2\,^2\!P$ &$-0.13024311\underline{963}$  &$-0.13023623\underline{876}$ &--0.130236\\
$3\,^2\!S$ &$-0.07418\underline{381350}$  &$-0.07417\underline{777025}$ &--0.074182\\
$3\,^2\!P$ &$-0.05723\underline{769823}$  &$-0.05723\underline{424577}$ &--0.057236\\
$3\,^2\!D$ &$-0.05561012\underline{974}$  &$-0.05560578\underline{543}$ &--0.055606\\
$4\,^2\!S$ &$-0.037\underline{52870957}$  &$-0.037\underline{52445073}$ &--0.038615\\
$4\,^2\!P$ &$-0.031\underline{39073613}$  &$-0.031\underline{38814390}$ &--0.031975\\
$4\,^2\!D$ &$-0.03127\underline{588444}$  &$-0.03127\underline{343938}$ &--0.031274\\
$4\,^2\!F$ &$-0.031253555\underline{31}$  &$-0.031251112\underline{02}$ &--0.031243\\
\hline
& \multicolumn{1}{c}{$^{\infty}$Be$^+$} & \multicolumn{1}{c}{$^{9}$Be$^+$} &  \multicolumn{1}{c}{$^{9}$Be$^+$} \\
$2\,^2\!S$ &$-0.66919693\underline{847}$    &$-0.66915422\underline{599}$ &--0.669247\\
$2\,^2\!P$ &$-0.52376705\underline{352}$    &$-0.52375065\underline{365}$ &--0.523769\\
$3\,^2\!S$ &$-0.2672\underline{0549176}$    &$-0.2671\underline{8867334}$ &--0.267233\\
$3\,^2\!P$ &$-0.2295\underline{6788615}$    &$-0.2295\underline{5822005}$ &--0.229582\\
$3\,^2\!D$ &$-0.22248781\underline{972}$    &$-0.22247429\underline{085}$ &--0.222478\\
$4\,^2\!S$ &$-0.13\underline{629487843}$    &$-0.13\underline{628082370}$ &--0.143152\\
$4\,^2\!P$ &$-0.12\underline{222924451}$    &$-0.12\underline{221999823}$ &--0.128134\\
$4\,^2\!D$ &$-0.12512\underline{688879}$    &$-0.12511\underline{926908}$ &--0.125124\\
$4\,^2\!F$ &$-0.1250154671\underline{1}$    &$-0.12500785\underline{769}$ &--0.125008\\
\end{tabular}
\end{ruledtabular}
\end{table}

\section{The structure calculations}
\subsection{Hamiltonian and Hylleraas coordinates}

The Li atom and Be$^+$ ion are four-body Coulomb systems. After
separating the center of mass coordinates, the nonrelativistic
Hamiltonian can be written in the form~\cite{zhang_yan}
\begin{eqnarray}
H_0 &=& -\sum_{i=1}^3 \frac{1}{2\mu}\nabla_i^2
-\frac{1}{m_0}\sum_{i> j\ge 1}^3\nabla_i\cdot\nabla_j
-\sum_{i=1}^3\frac{Z}{r_i} \nonumber \\
   & + &\sum_{i> j\ge 1}^3\frac{1}{r_{ij}}\,,
\label{eq:t1}
\end{eqnarray}
where $r_{ij}=|\textbf{r}_{i}- \textbf{r}_{j}|$ is the distance
between electrons $i$ and $j$, $\mu=m_0m_{\rm e}/(m_0+m_{\rm e})$ is
the reduced mass between the electron and the nucleus, and $Z$ is the
nuclear charge. In our calculation the wave functions are expanded
in terms of the explicitly correlated basis set in Hylleraas
coordinates:
\begin{eqnarray}
\phi({\bf{r}}_{1},{\bf{r}}_{2},{\bf{r}}_{3})&=&r_{1}^{j_{1}}r_{2}^{j_{2}}r_{3}^{j_{3}}r_{12}^{j_{12}}r_{23}^{j_{23}}r_{31}^{j_{31}}e^{-\alpha
r_{1}-\beta r_{2}-\gamma r_{3}} \nonumber \\
& \times &
\mathcal{Y}_{(\ell_{1}\ell_{2})\ell_{12},\ell_{3}}^{LM_L}(\hat{{\bf{r}}}_{1},\hat{{\bf{r}}}_{2},
\hat{{\bf{r}}}_{3})\chi(1,2,3)\,, \label{eq:t2}
\end{eqnarray}
where $\mathcal{Y}_{(\ell_{1}\ell_{2})\ell_{12},\ell_{3}}^{LM_L}$ is
the vector-coupled product of spherical harmonics to form an
eigenstate of total angular momentum $L$ and component $M_L$
\begin{eqnarray}
&&\mathcal{Y}_{(\ell_{1}\ell_{2})\ell_{12},\ell_{3}}^{LM_L}(\hat{{\bf{r}}}_{1},
\hat{{\bf{r}}}_{2},\hat{{\bf{r}}}_{3}) =\! \sum_{{\rm all\,}
m_{i}}\langle \ell_{1}m_{1}\ell_{2}m_{2}|\ell_{12}m_{12}\rangle
\nonumber\\
&&\times \langle\ell_{12}m_{12}\ell_{3}m_{3}|LM_L\rangle
Y_{\ell_{1}m_{1}}(\hat{{\bf{r}}}_{1})
Y_{\ell_{2}m_{2}}(\hat{{\bf{r}}}_{2})
Y_{\ell_{3}m_{3}}(\hat{{\bf{r}}}_{3})\,,\nonumber\\
 \label{eq:t3}
\end{eqnarray}
and $\chi(1,2,3)$ is the three-electron spin $1/2$ wave function.
The variational wave function is a linear combination of
anti-symmetrized basis functions $\phi$. With some truncations to
avoid potential numerical linear dependence, all terms in
Eq.~(\ref{eq:t2}) are included such that
\begin{eqnarray}
j_1+j_2+j_3+j_{12}+j_{23}+j_{31} \le \Omega\,, \label{eq:t4}
\end{eqnarray}
where $\Omega$ is an integer. The computational details in
evaluating the necessary matrix elements of the Hamiltonian may be
found in~\cite{yan-drake97}. The nonlinear parameters $\alpha$,
$\beta$, and $\gamma$ in Eq.~(\ref{eq:t2}) are optimized using
Newton's method.

The convergence for the energies and other expectation values is
studied by increasing $\Omega$ progressively. The basis sets are
essentially the same as two earlier Hylleraas calculations of the
static polarizabilities \cite{tang,tang2}. The maximum $\Omega$ used
in the present calculations is 12. The uncertainty in the final
value of any quantity is usually estimated to be equal to the size
of the extrapolation from the largest explicit calculation.

\begin{figure}
\includegraphics[width=0.49\textwidth]{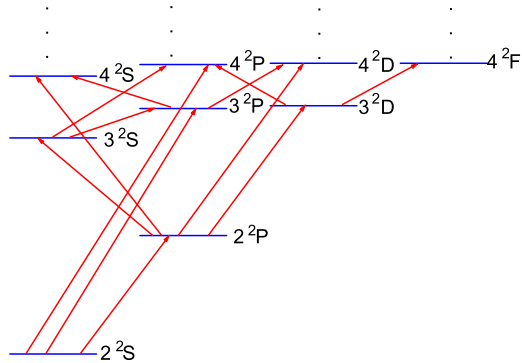}
\caption{Low lying energy levels of the Li atom. The
energy level diagram for Be$^+$ is similar. }
\label{f1}
\end{figure}

Fig.~\ref{f1} is a schematic diagram showing the nonrelativistic
energy levels of the most important states of the Li atom. The
energy level diagram for the low lying states of Be$^+$ is similar.

The energies of the ground states for $^\infty$Li and $^7$Li were
$-7.47806032391(5)$ and $-7.47745193065(5)$ a.u. respectively. The
respective energies for the $^\infty$Be$^+$ and $^9$Be$^+$ ground
states were $-14.3247631769(3)$ and $-14.3238634942(3)$ a.u.. Table
\ref{tab:1} gives the binding energies of the Li atom and Be$^+$ ion
systems with respect to the two-electron Li$^+$ and Be$^{2+}$ cores.
The Hylleraas basis was optimized to compute the $2\,^2\!S$ and
$2\,^2\!P$ state polarizabilities, so some of the $n = 4$ state
energies have significant deviations from the experimental $n = 4$
state energies. The states with significant energy differences can
be regarded as pseudo-states.  The uncertainties listed in Table
\ref{tab:1} represent the uncertainties in the energy with respect
to an infinite basis calculation. The actual computational
uncertainty is very small and there is no computational error in any
of the calculated digits listed in Table \ref{tab:1}.

With one exception, all the finite mass binding energies are less
tightly bound than experiment. The differences from experiment are
most likely due to relativistic effects. The exception where
experiment is less tightly bound than the finite mass calculation is
the $4\,^2\!F$ state of Li. This exception was not investigated
since the properties of this state do not enter into any of the
polarizability calculations.

\subsection{Polarizability definitions}

The dynamic polarizability provides a measure of the reaction of an
atom to an external electromagnetic field.  The dynamic polarizability
at real frequencies can be expressed in terms of a sum over all
intermediate states, including the continuum. The dynamic dipole
polarizability is expressed in terms of the dynamic scalar and tensor
dipole polarizabilities, $\alpha_1(\omega)$ and $\alpha_1^{T}(\omega)$,
which can be expressed in terms of the reduced matrix elements of
the dipole transition operator:
\begin{eqnarray}
\alpha_1(\omega) &=& \sum_{L_a} \alpha_1(L_a,\omega)\,, \label{eq:t5}\\
 \alpha_1^{T}(\omega) &=& \sum_{L_a}
W(L,L_a)\alpha_1(L_a,\omega)\,, \label{eq:t6}
\end{eqnarray}
where
\begin{eqnarray}
\alpha_1(L_a,\omega)=\frac{8\pi}{9(2L+1)}\sum_n\frac{\Delta
E_{0n}\big|\langle n_0L\|T_1\|nL_a\rangle\big|^2}{\Delta
E_{0n}^2-\omega^2} \,,\label{eq:t7}
\end{eqnarray}
with $T_1=\sum_{i=0}^3 q_iR_i Y_{10}({\bf {\hat{R}}}_i)$ being the
dipole transition operator, and
\begin{eqnarray}
W(L,L_a) &=& (-1)^{L+L_a}\sqrt{\frac{30(2L+1)L(2L-1)}{(2L+3)(L+1)}} \nonumber \\
& \times& \left\{ \begin{matrix}
  1 & 1 & 2 \\
  L & L & L_a \\
\end{matrix} \right\}\,.
 \label{eq:t8}
\end{eqnarray}
In the above, $|n_0L\rangle$ is the initial state with principal
quantum number $n_0$, angular momentum quantum number $L$, and
energy $E_0$. The $n$th intermediate eigenfunction $|nL_a\rangle$,
with principal quantum number $n$ and angular momentum quantum
number $L_a$, has an energy $E_n$. The transition energy is $\Delta
E_{0n}=E_n-E_0$. The $q_i$ are the charges of the respective
particles and ${\bf R}_i$ are defined in Ref.~\cite{zhang_yan}. In
particular, for the case of $L=0$,
\begin{eqnarray}
\alpha_1(\omega) &=& \alpha_1(P,\omega)\,, \label{eq:t9}\\
\alpha_1^{T}(\omega)&=& 0\,;
 \label{eq:t10}
\end{eqnarray}
for $L=1$,
\begin{eqnarray}
\alpha_1(\omega) \! &=& \! \alpha_1(S,\omega)+\alpha_1(P,\omega)+\alpha_1(D,\omega)\,, \label{eq:t11}\\
\alpha_1^{T}(\omega) \! &=& \! -\alpha_1(S,\omega)
+ \frac{1}{2}\alpha_1(P,\omega)-\frac{1}{10}\alpha_1(D,\omega).
 \label{eq:t12}
\end{eqnarray}
In  Eqs.~(\ref{eq:t11}) and (\ref{eq:t12}), $\alpha_1(P,\omega)$ is
the contribution from the even-parity configuration $(pp')P$. The
scalar and tensor polarizabilities can be easily related to the
polarizabilities of the magnetic sub-levels, $\alpha_{1,M}(\omega)$,
\begin{eqnarray}
\alpha_{1,0}(\omega)  &=& \alpha_1(\omega) - 2\alpha_1^{T}(\omega) \nonumber \\
\alpha_{1,\pm 1}(\omega)  &=& \alpha_1(\omega) + \alpha_1^{T}(\omega) \ .
\end{eqnarray}

\section{The dynamic polarizability for the $^{\infty}$Li atom and the $^{\infty}$Be$^+$ ion}

\subsection{Ground state dynamic polarizabilities}

\begin{figure}
\includegraphics[width=0.49\textwidth]{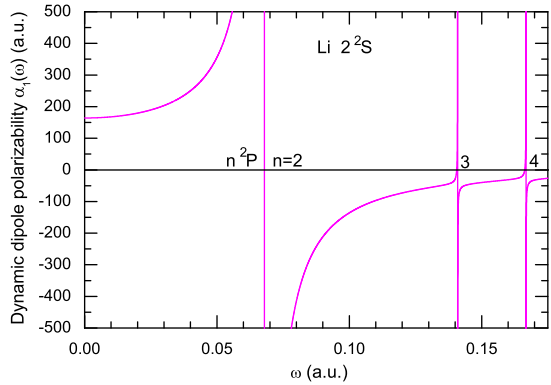}
\caption{Dynamic dipole polarizability, $\alpha_1(\omega)$, of the
Li atom in the ground state. The singularities in the polarizability
at the $2\,^2\!S \rightarrow n\,^2\!P$ frequencies are marked.}
\label{f2}
\end{figure}

Fig.~\ref{f2} shows the dynamic dipole polarizability of the lithium
ground state as a function of photon energy. The chief errors in the
dynamic polarizability are related to the convergence of the
$n\,^2\!P$ excited state energies. The largest calculation used a
basis with dimensions $(N_s,N_p)=(6412,5761)$. The difference
between the $\alpha_1(\omega)$ and polarizability computed with a
$(N_s,N_p)=(4172,3543)$ basis would be barely discernible in
Fig.~\ref{f2}. The convergence of $\alpha_1(\omega)$ is best at
photon energies far from the discrete excitation energies of the
$n\,^2\!P$ excitations. The polarizability is very susceptible to
small changes in the physical energies at photon energies close to
the $n\,^2\!P$ excitation energies.

\begingroup
\squeezetable
\begin{table*}[tbh]
\squeezetable \caption{The dynamic dipole polarizabilities,
$\alpha_1(\omega)$ (in a.u.), for the Li ground state. The results
of the fourth column include relativistic corrections. The numbers
in brackets for the second and third columns are the uncertainties
in the last digits arising from incomplete convergence of the basis
set. The uncertainties in the recommended (Rec.) values reflect
additional uncertainties related to the relativistic correction.}
\label{tab:2}
\begin{ruledtabular}
\begin{tabular}{lccccccc}
\multicolumn{1}{c}{$\omega$} & \multicolumn{3}{c}{Hylleraas} &
\multicolumn{1}{c}{MBPT-SD~\cite{safronova}} &
\multicolumn{1}{c}{TDGI} & \multicolumn{1}{c}{CI-Hylleraas } &
\multicolumn{1}{c}{Model }  \\
&  \multicolumn{1}{c}{$^{\infty}$Li} & \multicolumn{1}{c}{$^7$Li} &
\multicolumn{1}{c}{Rec. $^7$Li} &\multicolumn{1}{c}{} &
\multicolumn{1}{c}{~\cite{merawa94,merawa98}} &
\multicolumn{1}{c}{\cite{pipin}} &
\multicolumn{1}{c}{Potential~\cite{cohen}}  \\
\hline
0.00000  &   164.112(1) &164.161(1)  & 164.11(3)  &               & 163.6  & 164.1 & 164.14   \\
0.00500  &   164.996(1) &165.045(1)  & 165.00(3)  &               & 164.5  & 165.0 & 165.03   \\
0.01000  &   167.707(1) &167.758(1)  & 167.71(3)  &               & 167.2  & 167.7 & 167.74   \\
0.02000  &   179.517(1) &179.574(1)  & 179.52(3)  &               & 178.9  & 179.5 & 179.55   \\
0.02931  &  201.242(2)  &201.313(1)  & 201.24(3)  & 201.0(7)      &        &       &          \\
0.03000  &  203.438(1)  &203.512(1)  & 203.44(3)  &               & 202.6  & 203.4 & 203.47   \\
0.03420  &  219.221(1)  &219.307(1)  & 219.22(4)  & 219.0(8)      &        &       &          \\
0.04000  &  250.265(1)  &250.376(1)  & 250.26(4)  &               & 248.8  & 250.3 & 250.29   \\
0.04624  &  304.278(1)  &304.441(1)  & 304.26(5)  & 304.0(8)      &        &                 \\
0.05000  &  356.077(1)  &356.300(1)  & 356.05(6)  &               & 355.2  & 356.1 & 356.60   \\
0.05699  &  550.259(1)  &550.790(1)  & 550.18(9)  & 549.7(1.1)    &        &       &          \\
0.06000  &  741.165(2)  &742.126(1)  & 741.00(12)  &               & 729.2  &       & 740.73   \\
0.06507  &  1984.577(1) &1991.488(1) & 1983.11(31) & 1983(3)       &        &       &          \\
0.07000  & --2581.603(2)&--2569.994(2) & --2584.54(40) &               & --2895.3  &    &          \\
0.07592  & --645.478(2) &--644.749(1)  & --645.70(10)  & --645.9(1.3)  &           &    &          \\
0.08000  & --415.067(1) &--414.763(1)  & --415.17(7)  &               & --427.1   &    &          \\
0.09000  & --211.518(2) &--211.439(1)  & --211.55(3)  &               & --216.5   &    &          \\
0.09110  & --199.941(1) &--199.868(1)  & --199.97(3)  & --200.1(0.8)  &           &    &          \\
0.10000  & --135.872(2) &--135.838(1)  & --135.89(3)  &               & --0.819   &    &          \\
0.11388  & --86.266(1)  &--86.249(1)   & --86.27(2)   & --86.4(0.8)   &           &    &          \\
0.15183  & --38.210(9)  &--38.204(9)   & --38.22(1)   & --38.4(1.1)   &           &    &          \\
0.16000  & --31.08(5)   &--31.06(5)    & --31.08(6)    &    &           &    &          \\
\end{tabular}
\end{ruledtabular}
\end{table*}
\endgroup

The uncertainties in the dynamic dipole polarizabilities of the Li
ground state as well as the polarizabilities themselves are listed
in Table~\ref{tab:2}. All of the values listed are accurate to about
$\pm 1$ in the fifth digit for $\omega \leq 0.11388$ a.u.. Some of
the alternate calculations of the $\alpha_1(\omega)$
polarizabilities \cite{safronova,merawa94,merawa98,pipin,cohen} are
listed in Table~\ref{tab:2}. Dynamic polarizabilities from some less
accurate calculations \cite{chernov,kobayashi,muszynska} have not
been tabulated.

One feature of Table \ref{tab:2} is the excellent agreement with the
MBPT-SD calculation of Safronova {\it et al.}~\cite{safronova}. The
MBPT-SD calculation and the present Hylleraas calculation are in
perfect agreement when the MBPT-SD theoretical uncertainty is taken
into consideration. While the MBPT-SD calculation is fully
relativistic, its treatment of electron correlation is less exact
than the present calculation. The MBPT-SD calculation also gives no
consideration of finite mass effects. Relativistic effects would
tend to decrease $\alpha_1(\omega)$ at low $\omega$, and the MBPT-SD
calculation gives slightly smaller $\alpha_1(\omega)$ at low
$\omega$.

\begin{figure}
\includegraphics[width=0.49\textwidth]{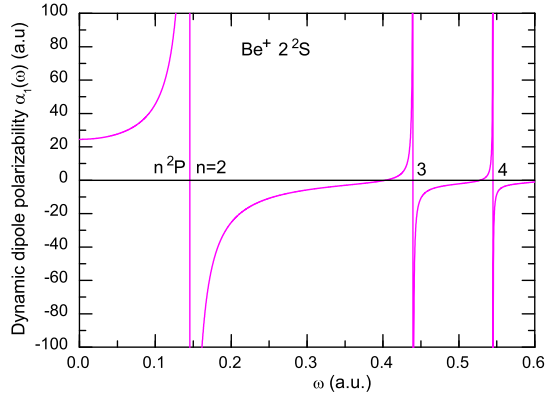}
\caption{The dynamic dipole polarizability, $\alpha_1(\omega)$ for
the ground state of the Be$^+$ ion. The singularities in the
polarizability at the $2\,^2\!S \rightarrow n\,^2\!P$ frequencies
are marked.} \label{f3}
\end{figure}

The older CI-Hylleraas calculation of values of Pipin and
Bishop~\cite{pipin} compares excellently with the present more
modern calculation. All digits in $\alpha_1(\omega)$ from the
CI-Hylleraas calculation are in perfect agreement with the present
Hylleraas calculation. The model potential polarizabilities of Cohen
and Themelis~\cite{cohen} are also very close to the present dynamic
polarizability. The Cohen-Themelis potential was constructed using a
Rydberg-Klein-Rees (RKR) inversion method. The
Time-Dependent-Gauge-Invariant (TDGI) polarizabilities of M\'{e}rawa
{\it et al}~\cite{merawa94,merawa98} are only accurate to 0.5$\%$ or
larger. The moderate accuracy of TDGI calculations has also been
noted in calculations of the static polarizabilities~\cite{tang}.

The static polarizabilities for Be$^+$ in the infinite mass
approximation have been presented recently~\cite{tang2}. The present
calculation represents an extension of this earlier calculation
since finite mass effects are now included. The dynamic dipole
polarizabilities listed in Table~\ref{tab:3} includes transition
frequencies that extend well into the ultraviolet region. The most
accurate of the few alternate calculations should be the
CI-Hylleraas calculation of Muszynska {\em et al} \cite{muszynska}.
However, it gives an $\alpha_1(\omega)$ that is about 1$\%$ smaller
than the present polarizability. Space limitations precluded
tabulation of the TDGI polarizability~\cite{merawa98}. The TDGI
polarizability was of only moderate accuracy with errors of about
$2\%$ for $\omega \leq 0.6$ a.u.. The uncertainty in the present
polarizability is about 10$^{-4}$ a.u. for photon energies lower
than 0.40 a.u., but has increased to 10$^{-2}$ a.u. at $\omega =
0.50$ a.u..

\begin{table}[tbh]
\caption{Dynamic dipole polarizabilities, $\alpha_1(\omega)$ (in
a.u.), for the ground state of the Be$^+$ ion. The results of the
fourth column incorporate relativistic effects. The numbers in
brackets are the uncertainties in the last digits arising from
incomplete convergence of the basis set. The recommended (Rec.)
polarizabilities in the fourth column reflect uncertainties other
than purely computational. } \label{tab:3}
\begin{ruledtabular}
\begin{tabular}{lcccc}
\multicolumn{1}{c}{$\omega$} & \multicolumn{3}{c}{Hylleraas} &
\multicolumn{1}{l}{CI-Hylleraas} \\
\multicolumn{1}{l}{(a.u.)} & \multicolumn{1}{c}{$^{\infty}$Be$^+$} &
\multicolumn{1}{c}{$^9$Be$^+$} & \multicolumn{1}{c}{Rec. $^9$Be$^+$}&
\multicolumn{1}{c}{~\cite{muszynska}}   \\ \hline
0.00 & 24.4966(1)    &24.5064(1)    & 24.489(4)   & 24.3 \\
0.01 & 24.6088(1)    &24.6187(1)    & 24.601(4)   & 24.4 \\
0.02 & 24.9518(1)    &24.9620(1)    & 24.943(4)   & 24.7 \\
0.04 & 26.4291(1)    &26.4404(1)    & 26.419(4)   & 26.2 \\
0.06 & 29.3390(1)    &29.3528(1)    & 29.325(5)   & 29.1 \\
0.08 & 34.7358(1)    &34.7550(1)    & 34.715(6)   & 34.3 \\
0.10 & 45.6509(1)    &45.6836(1)    & 45.609(7)   & 44.9 \\
0.12 & 74.7857(1)    &74.8724(1)    & 74.656(12)   &      \\
0.15 & $-$367.8708(2)&$-$365.8030(3)& $-$371.860(60)&               \\
0.18 & $-$43.2038(1) &$-$43.1745(1) & $-43.273(7)$ &               \\
0.20 & $-$25.3195(1) &$-$25.3090(1) & $-25.348(4)$ &              \\
0.30 & $-$5.7967(1)  &$-$5.7951(1)  & $-5.801(2)$ &               \\
0.40 & $-$0.2912(1)  &$-$0.2873(1)  & $-0.2961(2)$ &              \\
0.50 & $-$2.149(7)   &$-$2.161(7)   & $-2.164(8)$ &              \\
\end{tabular}
\end{ruledtabular}
\end{table}

The dynamic polarizabilities for the Li and Be$^+$ ground
states are depicted in Figures \ref{f2} and \ref{f3}. There are
obvious similarities in shapes of the two $\alpha_1(\omega)$ curves
but with the Li polarizability being about 5-10 times larger in
magnitude at comparable values of $\omega/\omega_{2s \to 2p}$. One
difference between the two curves is that Be$^+$ has zeroes
in $\alpha_1({\omega})$ at a discernible frequency difference before
the $3\,^2\!P$ and $4\,^2\!P$ excitations while the
$\alpha_1(\omega)$ negative to positive crossovers for Li
occur much closer to the transition frequencies.

\subsection{Excited state dynamic polarizabilities}

\begin{figure}
\includegraphics[width=0.49\textwidth]{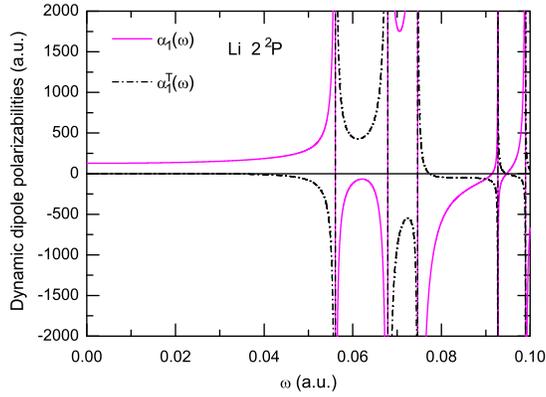}
\caption{The dynamic polarizabilities, $\alpha_1(\omega)$ and
$\alpha_1^T(\omega)$ (in a.u.) of the Li $2\,^2\!P$ state for photon
frequencies below 0.10 a.u.. The scalar polarizability is given by
the solid line while the tensor polarizability is given by the chain
curve.} \label{f4}
\end{figure}

The scalar and tensor dipole polarizabilities for the excited
$2\,^2\!P$ state of the Li atom are listed in Table \ref{tab:4}. As
far as we know the present calculations are the only dynamic
polarizabilities presented for this state. The structure of the
dynamic polarizability is complicated since both downward and upward
transitions leads to singularities. This is seen most clearly in
Fig.~\ref{f4} which plots the polarizabilities for photon energies
up to 0.10 a.u.. The tensor polarizability is generally small except
in the vicinity of the $2\,^2\!S$, $3\,^2\!S$ and $3\,^2\!D$
transitions. The tensor polarizability can become large when a
single transition tends to dominate Eq.~(\ref{eq:t12}). The scalar
and tensor polarizabilities tend to be opposite in sign. The main
contribution to the polarizabilities comes from transitions to the
$S$ and $D$ states. The coefficients in the sum-rules,
Eqs.~(\ref{eq:t11}) and (\ref{eq:t12}), for these terms are opposite
in sign.

\begin{table*}[tbh]
\caption{The dynamic dipole polarizabilities of the $2\,^2\!P$ state
of Li and Be$^+$. Both the scalar and tensor polarizabilities are
tabulated. The numbers in brackets are the uncertainties in the last
digits arising from incomplete convergence of the basis set. Values
without uncertainties have no numerical uncertainties in any of the
quoted digits. The recommended (Rec.) polarizabilities in the sixth
and seventh columns have estimated corrections from relativistic
effects. The recommended polarizabilities reflect uncertainties
other than purely computational.} \label{tab:4}
\begin{ruledtabular}
\begin{tabular}{lrrrrrr}
\multicolumn{1}{c}{$\omega$} & \multicolumn{2}{c}{$^{\infty}$Li} & \multicolumn{2}{c}{$^{7}$Li} & \multicolumn{2}{c}{Rec. $^{7}$Li} \\
                        \cline{2-3} \cline{4-5}\cline{6-7}
(a.u.)    & \multicolumn{1}{c}{$\alpha_1$} &
\multicolumn{1}{c}{$\alpha_T$} & \multicolumn{1}{c}{$\alpha_1$} &
\multicolumn{1}{c}{$\alpha_T$}  & \multicolumn{1}{c}{$\alpha_1$} &
\multicolumn{1}{c}{$\alpha_T$} \\ \hline
0.00 & 126.9458(3)    & 1.6214(3)    & 126.9472(5)    & 1.6351(2)    & 126.970(4)   & 1.612(4)\\
0.01 & 129.2491(5)    & 1.4035(2)    & 129.2501(5)    & 1.4178(2)    & 129.273(4)   & 1.393(4)\\
0.02 & 136.8371(5)    & 0.5302(3)    & 136.8372(5)    & 0.5463(5)    & 136.864(4)   & 0.518(4)\\
0.03 & 152.469(1)     & $-$2.091(1)  & 152.468(2)     & $-$2.070(1)  & 152.503(4)   & $-2.106(4)$\\
0.04 & 185.542(5)     & $-$11.722(5) & 185.535(5)     & $-$11.691(5) & 185.593(6)   & $-11.747(6)$\\
0.05 & 301.24(8)      & $-$82.33(9)  & 301.23(9)      & $-$82.27(8)  & 301.33(10)    & $-82.38(10)$\\
0.06 & $-$119.1(5)    & 446.7(5)     & $-$119.3(5)    & 446.9(5)     & $-119.0(6)$  & 446.6(6)\\
0.07 & 1804.5(1)      & $-$904.2(2)  & 1801.2(1)      & $-$900.34(5) & 1806.2(2)    & $-905.2(2)$\\
0.08 & $-$593.1(2)    & $-$43.5(4)   & $-$592.7(3)    & $-$43.6(5)   & $-592.6(5)$  & $-43.4(5)$\\
\hline
\multicolumn{1}{l}{} & \multicolumn{2}{c}{$^{\infty}$Be$^+$} & \multicolumn{2}{c}{$^9$Be$^+$}  & \multicolumn{2}{c}{Rec. $^9$Be$^+$} \\
                        \cline{2-3} \cline{4-5}\cline{6-7}
                             & \multicolumn{1}{c}{$\alpha_1$} & \multicolumn{1}{c}{$\alpha_T$} & \multicolumn{1}{c}{$\alpha_1$} & \multicolumn{1}{c}{$\alpha_T$}
                             & \multicolumn{1}{c}{$\alpha_1$} & \multicolumn{1}{c}{$\alpha_T$}  \\ \hline
0.00 & 2.02476(1)    &5.856012(1)     &2.02319(1)       &5.858938(1)    &2.0285(10)    &5.8528(10)\\
0.01 & 1.99755(1)    &5.890887(1)     &1.99595(1)       &5.893842(1)    &2.0013(10)    &5.8876(10)\\
0.02 & 1.91389(1)    &5.997630(1)     &1.91221(1)       &6.000672(1)    &1.9178(12)     &5.9942(12)\\
0.05 & 1.23144(1)    &6.845876(1)     &1.22905(1)       &6.849643(1)    &1.2363(13)    &6.8415(13)\\
0.10 & $-$3.88178(1) &12.61790    &$-$3.89080(1)    &12.62842           &$-3.8666(15)$  &12.6033(13)\\
0.14 & $-$94.71454   &104.48595  &$-$95.2476   &105.02073               &$-93.7873(16)$ &103.5592(16)\\
0.20 & 26.1505(1)     &$-$12.9481(2)  &26.1499(1)       &$-$12.9451(1)  &26.161(17)     &$-12.958(12)$\\
0.25 & 48.59(5)       &$-$26.10(1)    &48.61(5)         &$-$26.11(1)    &48.62(6)      &$-26.17(1)$\\
0.28 & 52.09(1)       &$-$3.43(1)     &52.10(1)         &$-$3.44(1)     &52.07(1)      &$-3.45(1)$\\
0.32 & $-$49.87(1)    &4.413(1)       &$-$49.85(1)      &4.412(1)       &$-49.89(1)$   &4.40(1)\\
\end{tabular}
\end{ruledtabular}
\end{table*}

\begin{figure}
\includegraphics[width=0.49\textwidth]{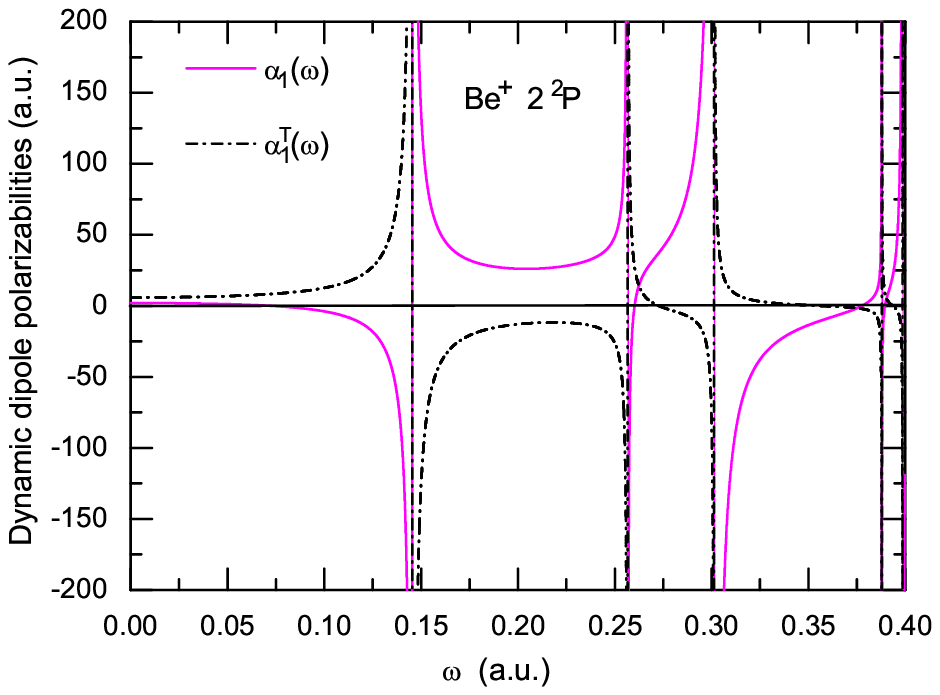}
\caption{The dynamic polarizabilities, $\alpha_1(\omega)$ and
$\alpha_1^T(\omega)$ (in a.u.), of the Be$^+$ $2\,^2\!P$ state for
photon frequencies below 0.40 a.u..  The scalar polarizability is
given by the solid line while the tensor polarizability is given by
the chain curve.} \label{f5}
\end{figure}

\begin{figure}
\includegraphics[width=0.49\textwidth]{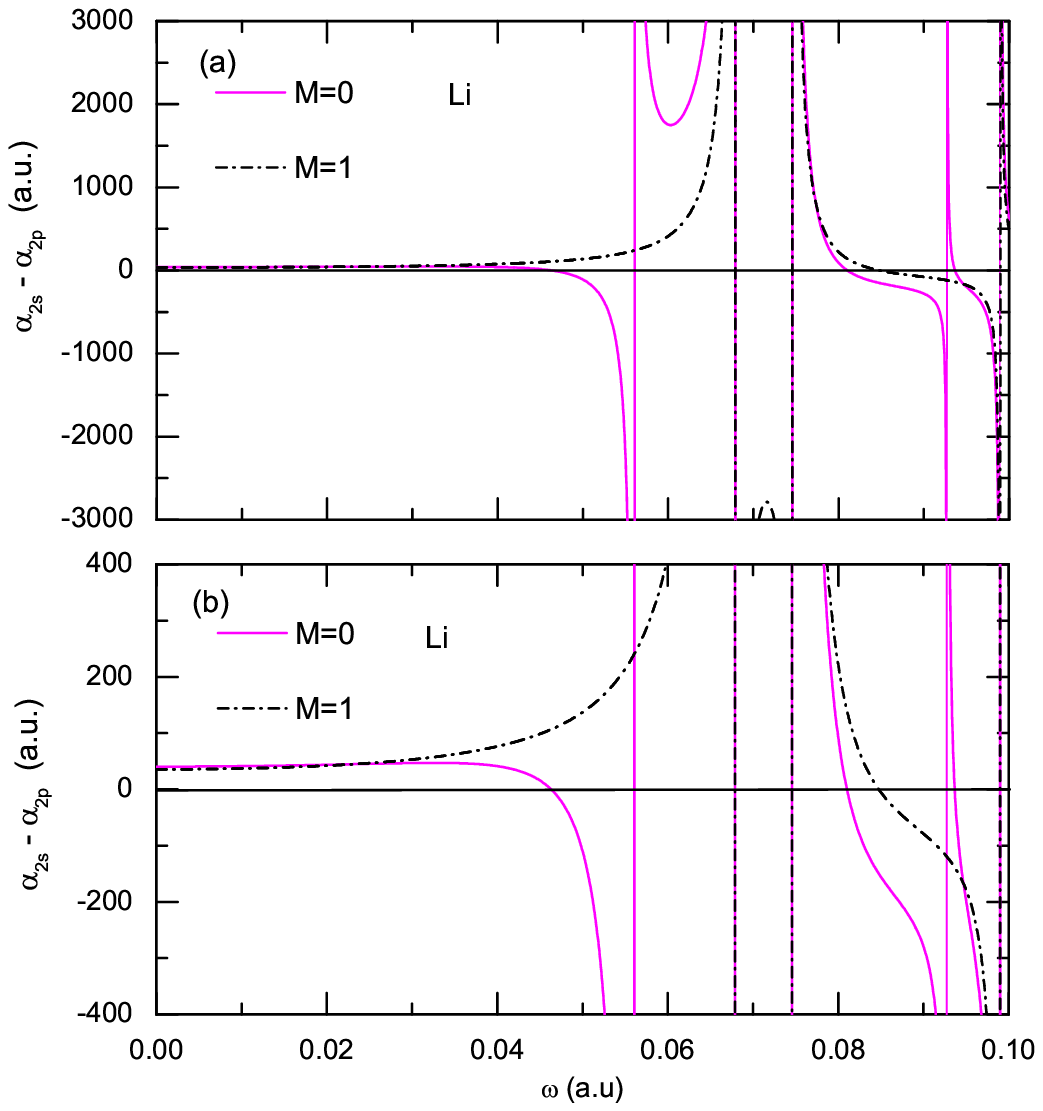}
\caption{The polarizability difference between the $2\,^2\!S$ and
$2\,^2\!P$ states of Li. Polarizability differences are shown
for $M = 0$ and $M = 1$.} \label{f6}
\end{figure}

The dynamic polarizabilities for the Be$^+$ $2\,^2\!P$ state are
also tabulated in Table \ref{tab:4} and depicted in Figure \ref{f5}
for the photon frequencies below 0.40 a.u.. There are three
resonances in this frequency range. The scalar and tensor dynamic
polarizabilities are similar in shape but with the opposite sign. As
far as we know, there has been no previous calculation of the
$2\,^2\!P$ state dynamic polarizability.

\begin{figure}
\includegraphics[width=0.49\textwidth]{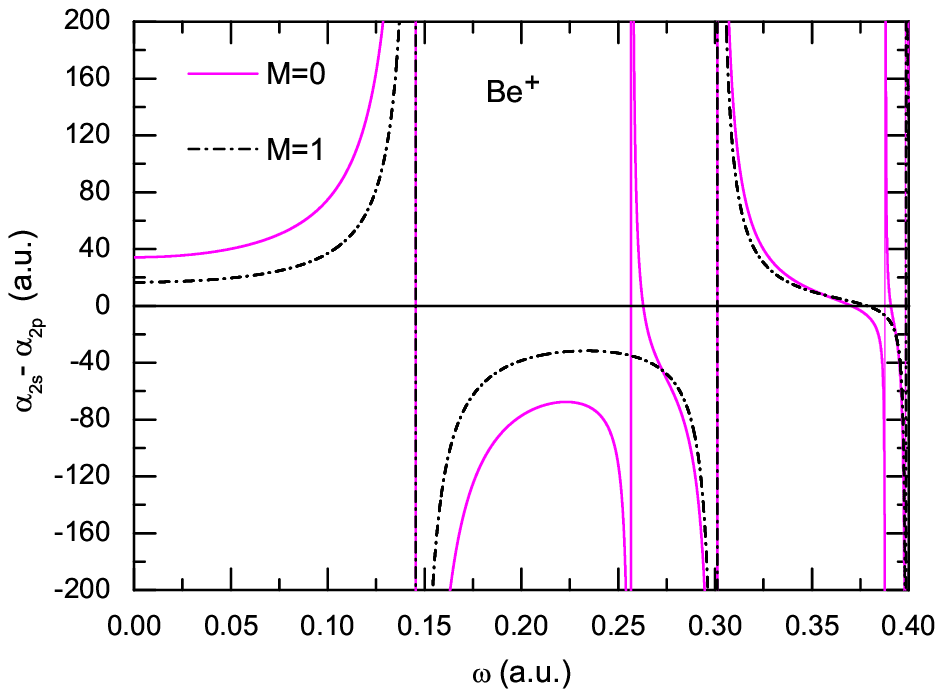}
\caption{The polarizability difference between the $2\,^2\!S$ and
$2\,^2\!P$ states of Be$^+$. Polarizability differences are
shown for $M = 0$ and $M = 1$. } \label{f7}
\end{figure}

\subsection{The static $2\,^2\!S$ $\to$ $2\,^2\!P$ Stark shift}

The static Stark shift for the $2\,^2\!S$ $\to$ $2\,^2\!P$ energy
interval in an electric field of strength $F$ is written as
\begin{eqnarray}
\Delta E_{\rm 2s-2p, M} &=& -\frac{1}{2} F^2 \left(
                     \alpha_{\rm 2s}  - \alpha_{\rm 2p, M} \right) \nonumber \\
                   &  - & \frac{1}{24} F^4
           \left( \gamma_{\rm 2s} - \gamma_{\rm 2p, M} \right)
           + \ldots\,,\label{eq:t19}
\end{eqnarray}
where $\gamma$ is the hyper-polarizability. The Stark shift depends
on the magnetic quantum number $M$ of the $2\,^2\!P$ state. The
relative size of $\Delta \alpha$ and $\Delta \gamma$ determines the
extent to which the Stark shift is influenced by the
hyper-polarizability at high field strengths. The relative
importance of $\Delta \alpha$ and $\Delta \gamma$ is given by the
ratio
\begin{equation}
X=\frac{F^2(\gamma_{2s}-\gamma_{2p, M})}{12(\alpha_{2s}-\alpha_{2p,
M})}=\frac{F^2\Delta \gamma}{12\Delta \alpha}\,.\label{eq:t20}
\end{equation}
Using the static polarizability and static hyper-polarizability for
the Li atom results in $\Delta \alpha = 37.1$ and $\Delta \gamma =
9.99 \times 10^6$ giving $X = 0.0001$ at $F = 6.67 \times 10^{-5}$
a.u. (344 kV/cm) and $X=0.001$ at $F=2.11\times 10^{-4}$ a.u. (1087
kV/cm). These estimates of the critical field strength where the
quadratic Stark shift is valid depend slightly on the magnetic
quantum number and exact values can be determined by using
$M$-dependent polarizabilities. Stark shifts of higher order than
the hyper-polarizability can be comfortably ignored at the 0.01$\%$
level provided the field strength is less than 1100 kV/cm. The
static Stark shift for Be$^+$ is not interesting since it is
difficult to measure as a Be$^+$ ion immersed in a finite electric
field is accelerated away from the finite field region.

\begin{table}[tbh]
\caption{The photon energies for which there is no Stark shift for
the $2\,^2\!S$ $\to$ $2\,^2\!P$ transition. Underlined digits
indicate uncertain digits arising from lack of basis set
convergence. Digits in brackets indicate possible uncertainties
associated with relativistic corrections in the recommended (Rec.)
values.} \label{tab:5}
\begin{ruledtabular}
\begin{tabular}{lcc}
\multicolumn{1}{c}{System} & \multicolumn{1}{c}{$M=0$} & \multicolumn{1}{c}{$M=1$}  \\ \hline
$^\infty$Li     & $0.046317680\underline{6}$     & $0.084763957\underline{1}$\\
                & $0.081021795\underline{5}$     & \\
                & $0.093664330\underline{5}$     & \\
$^7$Li          & $0.046335687\underline{8}$     & $0.084766087\underline{0}$\\
                & $0.081024478\underline{9}$     & \\
                & $0.093661899\underline{1}$     & \\
Rec. $^7$Li & $0.046297(4)$     & $0.084756(2)$\\
                & $0.081014(2)$     & \\
                & $0.0936613(2)$     & \\
$^\infty$Be$^+$ & $0.262920267\underline{8}$     & $0.378457000\underline{4}$\\
                & $0.370371502\underline{7}$     &\\
                & $0.390752146\underline{3}$     &\\
$^9$Be$^+$      & $0.262917360\underline{3}$     & $0.378451843\underline{7}$\\
                & $0.370279952\underline{2}$     &\\
                & $0.390455356\underline{8}$     &\\
Rec. $^9$Be$^+$  & $0.2628956(7)$     & $0.378443(2)$\\
                & $0.370274(1)$     &\\
                & $0.3904546(2)$     &\\
\end{tabular}
\end{ruledtabular}
\end{table}

\subsection{The dynamic $2\,^2\!S$ $\to$ $2\,^2\!P$ Stark shift }

The Li Stark shifts, $\alpha(2s)- \alpha(2p_M)$, are plotted as a
function of frequency in Figure~\ref{f6}. It is seen that there are
magic wavelengths for $M = 0$ just below the $2\,^2\!P$ $\to$
$3\,^2\!S$ threshold and between the $2\,^2\!S$ $\to$ $2\,^2\!P$ and
$2\,^2\!P$ $\to$ $3\,^2\!D$ thresholds. The actual energies for
which the polarizability difference is zero are given in Table
\ref{tab:5}. The Stark shifts get very large for frequencies between
0.058 and 0.070 a.u..

The Be$^+$ Stark shifts, $\alpha(2s)- \alpha(2p_M)$, are plotted as
a function of frequency in Figure~\ref{f7}. The Stark shifts are
much smaller in magnitude than the Li atom shifts. One difference
from Li is that the Be$^+$ shift has no zero for energies below the
$2\,^2\!S$ $\to$ $2\,^2\!P$ threshold. The first zero in the Stark
shift (excepting those related to a singularity) is at 0.263 a.u..

\begingroup
\squeezetable
\begin{table*}[tbh]
\caption{The parameters for the calculation of the $2\,^2\!S$ state
frequency-dependent polarizabilities of Li and Be$^+$. The numbers
in the square brackets denote powers of 10. The recommended (Rec.)
results of the fourth and the seventh columns incorporate
relativistic corrections.} \label{tab:6}
\begin{ruledtabular}
\begin{tabular}{lcccccc}
\multicolumn{1}{l}{Parameter} & \multicolumn{1}{c}{$^{\infty}$Li}&
\multicolumn{1}{c}{$^7$Li} & \multicolumn{1}{c}{Rec. $^7$Li}
                              &  \multicolumn{1}{c}{$^{\infty}$Be$^+$}& \multicolumn{1}{c}{$^9$Be$^+$}
                              & \multicolumn{1}{c}{Rec. $^9$Be$^+$}\\
\hline
$f_{2s\rightarrow 2p}$  &0.746956$\underline{855381}$    &0.746961$\underline{871867}$       &0.747011$\underline{776131}$    &0.4980674$\underline{22721}$    &0.4980833$\underline{82699}$  &0.4982270$\underline{10322}$\\
$\Delta E_{2s2p}$       &0.06790379$\underline{1567}$    &0.06789417$\underline{2078}$       &0.06790605                      &0.14542988$\underline{4364}$    &0.14540357$\underline{2344}$  &0.14547806\\
$f_{2s\rightarrow 3p}$  &0.00473$\underline{1019443}$    &0.00473$\underline{7600312}$       &0.00472$\underline{8028090}$    &0.0832$\underline{43986131}$    &0.0832$\underline{889414647}$  &0.0832$\underline{09271939}$\\
$\Delta E_{2s3p}$       &0.14090$\underline{9212964}$    &0.14089$\underline{6165068}$       &0.14090640                      &0.4396$\underline{29051730}$    &0.4395$\underline{96005937}$  &0.43966521\\
$f_{2s\rightarrow 4p}$  &0.00496$\underline{0028680}$    &0.00496$\underline{4714658}$       &                                &0.04$\underline{0874056901}$    &0.04$\underline{0896543934}$  &\\
$\Delta E_{2s4p}$       &0.166$\underline{756175058}$    &0.166$\underline{742266938}$       &                                &0.54$\underline{6967693380}$    &0.54$\underline{6934227760}$  &\\
$S(-2)$                 &1.697$\underline{71}$           &1.698$\underline{63}$              &                                &0.37$\underline{9627}$          &0.37$\underline{9781}$  &\\
$S(-4)$                 &21.$\underline{8714}$           &21.$\underline{8888}$              &                                &0.5$\underline{13938}$          &0.5$\underline{14207}$  &\\
$S(-6)$                 &4$ \underline{28.809}$          &4$\underline{29.237}$              &                                &0.9$\underline{79476}$          &0.9$\underline{80088}$  &\\
$S(-8)$                 &9.$\underline{69364}[3]$        &9.$\underline{70502}[3]$           &                                &2.$\underline{05322}$           &2.$\underline{05469}$  &\\
$S(-10)$                &2.$\underline{37008}[5]$        &2.$\underline{37325}[5]$           &                                &4.$\underline{52145}$           &4.$\underline{52508}$  &\\
$S(-12)$                &6.$\underline{08098}[6]$        &6.$\underline{09006}[6]$           &                                &1$\underline{0.2473}$           &1$\underline{0.2564}$  &\\
$S(-14)$                &1.$\underline{61032}[8]$        &1.$\underline{61296}[8]$           &                                &2$\underline{3.6368}$           &2$\underline{3.6596}$  &\\
$S(-16)$                &4.$\underline{35760}[9]$        &4.$\underline{36537}[9]$           &                                &5$\underline{5.1280}$           &5$\underline{5.1852}$  &\\
$\eta_1$                &2$\underline{7.0605}$           &2$\underline{7.0643}$              &                                &2.$\underline{33229}$           &2.$\underline{33246}$  &\\
\end{tabular}
\end{ruledtabular}
\end{table*}
\endgroup

\subsection{Analytic representation}

The utility of the present calculations can be increased by
constructing a closed form expression for the dynamic
polarizability. This is done by retaining the first $3$ terms in
Eq.~(\ref{eq:t7}) explicitly and then expanding the energy
denominator in the remainder. The expressions explicitly include
oscillator strengths up to the $n = 4$ principal quantum numbers.
The closed form expression is
\begin{eqnarray}
\alpha_1(\omega)& = & \left(  \sum_{n=2}^{4}
\frac{f_{2s \rightarrow np}}{\Delta E_{2snp}^2-\omega^2} \right)
 + S(-2) + \omega^2 S(-4) \nonumber \\
 &+& \omega^4 S(-6) + \ldots +\omega^{14} S(-16) + C(\omega)
\label{eq:t13}
\end{eqnarray}
where
\begin{equation}
S(-m)=\sum_{n=5} \frac{f_{2s \rightarrow np}}{(\Delta E_{2snp})^{m}}
\,,\label{eq:t14}
\end{equation}
\begin{equation}
C(\omega)=\frac{\eta_1 \omega^{16}S(-16)}{1-\eta_1\omega^2} \,.
\end{equation}
Here $f_{2s \rightarrow np}$ are the dipole oscillator strengths for
the $2\,^2\!S \rightarrow n\,^2\!P$ transitions with transition
energies $\Delta E_{2snp}$. The $S(-n)$ are the Cauchy moments of
the remainder of the oscillator strength distribution and are
independent of $\omega$. The $C(\omega)$ is an approximate term to
represent the summation from the term $S(-18)$ to $S(\infty)$. The
ratio, $\eta_1=S(-n-2)/S(-n)$, is assumed to be constant and its
value is set at $S(-16)/S(-14)$. Numerical values of the various
constants in Eq.~(\ref{eq:t13}) can be found in Table \ref{tab:6}.
Inclusion of the remainder term has greatly increased the precision
of the analytic fit to the exact dynamic polarizability.

The analytic representation for the Li $2\,^2\!S$ state is
accurate to 0.01 a.u. for $\omega \leqslant 0.1612$ a.u. and to an
accuracy of 0.1 a.u. for $\omega \leqslant 0.1728$ a.u.. The dynamic
polarizability for the Be$^+$ $2\,^2\!S$ state maintains its accuracy
over a larger $\omega$ range. It is accurate to 0.001 a.u. for
$\omega \leqslant 0.543$ a.u., to 0.01 a.u. for $\omega \leqslant
0.58605$ a.u. and 0.1 a.u. for $\omega \leqslant 0.6$ a.u..

The presence of zeroes in the dynamic polarizability near the
singularities means that the relative error in the analytic
representation can get very large in a frequency range very close to
the zeroes. Neglecting these localized regions with anomalously high
relative uncertainties, the relative difference between the analytic
representation and actual dynamic polarizability for the Li
$2\,^2\!S$ state was less than 0.001\% for $\omega \leqslant 0.1399$
a.u., 0.01\% for $\omega \leqslant 0.1551$ a.u., and 0.1\% for
$\omega \leqslant 0.1651$ a.u.. The relative difference for the
Be$^+$ $2\,^2\!S$ state obtained by the variational Hylleraas method
was less than 0.001\% for $\omega \leqslant 0.4737$ a.u., 0.01\% for
$\omega \leqslant 0.5067$ a.u. and 0.1\% for $\omega \leqslant
0.52575$ a.u.. The inclusion of the remainder term, $C(\omega)$
improved the accuracy of the analytic representation by one or two
order of magnitude within the frequency range listed above.

The dynamic dipole polarizabilities of the $2\,^2\!P$ states of Li
and Be$^+$ have both scalar, $\alpha_1(\omega)$, and tensor,
$\alpha_1^{T}(\omega)$, parts. The scalar part can be written
\begin{eqnarray}
\alpha_1(\omega) &=&   \sum_{n=2}^{4}
\frac{f_{2p \rightarrow ns}}{\Delta E_{2pns}^2-\omega^2}
 +  \sum_{n=3}^{4} \frac{f_{2p \rightarrow nd}}{\Delta E_{2pnd}^2-\omega^2} \nonumber \\
 &+& S(-2) + \omega^2 S(-4) + \omega^4 S(-6) +  \ldots  \nonumber \\
 &+& \omega^{14}S(-16)+C(\omega)\,,
\label{eq:t15}
\end{eqnarray}
where
\begin{eqnarray}
S(-m)&=&\sum_{n=5} \frac{f_{2p \rightarrow ns}}{(\Delta E_{2pns})^{m}}
 +  \sum_{n'} \frac{f_{2p \rightarrow n'P}}{(\Delta E_{2pn'P})^{m}} \nonumber \\
 &+&  \sum_{n=5} \frac{f_{2p \rightarrow nd}}{(\Delta E_{2pnd})^{m}} \,.\label{eq:t16}
\end{eqnarray}

The $2p \rightarrow n'P$ excitation involves a core excitation and
the intermediate state is an unnatural parity $^2\!P^e$ state. The
tensor part is
\begin{eqnarray}
\alpha^T_1(\omega) &= & -\sum_{n=2}^{4}
\frac{f_{2p \rightarrow ns}}{\Delta E_{2pns}^2-\omega^2}
 - \frac{1}{10} \sum_{n=3}^{4}
\frac{f_{2p \rightarrow nd}}{\Delta E_{2pnd}^2-\omega^2} \nonumber \\
 &+& S^T(-2) + \omega^2 S^T(-4) + \omega^4 S^T(-6) + \ldots
 \nonumber\\
 &+& \omega^{14}S^T(-16)+C^T(\omega)\,,
\label{eq:t17}
\end{eqnarray}
where
\begin{eqnarray}
S^T(-m) & = & -\sum_{n=5} \frac{f_{2p \rightarrow ns}}{(\Delta E_{2pns})^{m}}
 +  \frac{1}{2} \sum_{n'} \frac{f_{2p \rightarrow n'P}}{(\Delta E_{2pn'P})^{m}} \nonumber \\
 &-&  \frac{1}{10} \sum_{n=5} \frac{f_{2p \rightarrow nd}}{(\Delta E_{2pnd})^{m}} \,,\label{eq:t18}
\end{eqnarray}
\begin{eqnarray}
C^T(\omega)=\frac{\eta^T_1\omega^{16} S^T(-16)}{1-\eta^T_1\omega^2} \,,
\end{eqnarray}
where $f_{2p\rightarrow mL_1}$ means the oscillator strength from
$2p$ state to $mL_1$ state transition. $S^T(-2)$, $S^T(-4)$,
$S^T(-6)$ $\cdots$ are the coefficients corresponding to $\omega^0$,
$\omega^2$, $\omega^4$ $\cdots$ terms of the tensor part. The
remainder term, $C^T(\omega)$ is an approximate expression to
take into account the $S^T(-18)$ $\to$ $S^T(\infty)$ summations.
The factor $\eta^T_1$ is set to be $\eta^T_1 = S(-16)/S(-14)$.
All parameters in the analytic representation
are given in Table~\ref{tab:7}.

The first two terms of Eqs.~(\ref{eq:t15}) and (\ref{eq:t17})
include five resonances, which make the major contribution to the
polarizability with the second term involving excitations to $D$
states being the most important. This is clearly seen in the Li
$\alpha_1(\omega)$ of 185.542(5) a.u. at $\omega=0.04$ a.u.. The
contribution of the first summation of Eq.~(\ref{eq:t15}) was
$-8.8717$ a.u., while the second summation contributed 175.8241
a.u.. The value given by Eq.~(\ref{eq:t15}) was 185.5378 a.u., which
agrees with the exact value at the level of 0.0004$\%$.

The analytic representation for the scalar polarizability
$\alpha_1(\omega)$ of the Li $2\,^2\!P$ state is accurate to 0.01
a.u. for $\omega \leqslant 0.0855$ a.u. and to 0.1 a.u. for $\omega
\leqslant 0.0937$ a.u.. The analytic representation for the tensor
polarizability, $\alpha_1^T(\omega)$, is accurate to 0.01 a.u. for
$\omega \leqslant 0.0926$ a.u. and to 0.10 a.u. for $\omega
\leqslant 0.1$ a.u..

\begingroup
\squeezetable
\begin{table*}[tbh]
\caption{The parameters defining the frequency-dependent
polarizabilities of the $2\,^2\!P$ state of Li and Be$^+$. The
numbers in the square brackets denote powers of 10. The recommended
(Rec.) results of the fourth and the seventh columns incorporate
relativistic effects.} \label{tab:7}
\begin{ruledtabular}
\begin{tabular}{lccccccc}
 \multicolumn{1}{l}{Parameter} & \multicolumn{1}{c}{$^{\infty}$Li}& \multicolumn{1}{c}{$^7$Li}
 & \multicolumn{1}{c}{Rec. $^7$Li} & \multicolumn{1}{c}{$^{\infty}$Be$^+$} & \multicolumn{1}{c}{$^9$Be$^+$}
 & \multicolumn{1}{c}{Rec. $^9$Be$^+$}\\ \hline
$f_{2p\rightarrow 2s}$  &$-0.2489856\underline{18454}$        &$-0.2489872\underline{90622}$     &$-0.2490039\underline{2538}$    &$-0.1660224\underline{74240}$   &$-0.1660277\underline{94233}$    &$-0.1660756\underline{70107}$\\
$\Delta E_{2p2s}$       &$-0.06790379\underline{1567}$        &$-0.06789417\underline{2078}$     &$-0.067906050$                  &$-0.14542988\underline{4364}$   &$-0.14540357\underline{2344}$    &$-0.145478060000$\\
$f_{2p\rightarrow 3s}$  &0.1105$\underline{78835460}$         &0.1105$\underline{75403831}$      &0.1105$\underline{89306872}$    &0.0643$\underline{85095804}$    &0.0643$\underline{85347515}$     &0.0644$\underline{07168594}$\\
$\Delta E_{2p3s}$       &0.05605$\underline{9306121}$         &0.05605$\underline{8468507}$      &0.056054150                     &0.25656$\underline{1561765}$    &0.25656$\underline{1980308}$     &0.256536125000\\
$f_{2p\rightarrow 4s}$  &0.014$\underline{979087136}$         &0.014$\underline{981586569}$      &                                &0.012$\underline{876689561}$    &0.012$\underline{879003186}$     &\\
$\Delta E_{2p4s}$       &0.0927$\underline{14410055}$         &0.0927$\underline{11788029}$      &                                &0.387$\underline{472175091}$    &0.387$\underline{469829948}$     &\\
$f_{2p\rightarrow 3d}$  &0.638568$\underline{044661}$         &0.638583$\underline{007678}$      &0.638583$\underline{083728}$    &0.63198$\underline{1700709}$    &0.63205$\underline{9480294}$     &0.63204$\underline{7210702}$\\
$\Delta E_{2p3d}$       &0.07463298$\underline{9884}$         &0.07463045$\underline{3329}$      &0.074630150                     &0.3012792$\underline{33806}$    &0.3012763$\underline{62793}$     &0.301291440000\\
$f_{2p\rightarrow 4d}$  &0.1227$\underline{46501135}$         &0.1227$\underline{56411337}$      &                                &0.12271$\underline{1398708}$    &0.12273$\underline{4598764}$\\
$\Delta E_{2p4d}$       &0.098967$\underline{235189}$         &0.098962$\underline{799378}$      &                                &0.39864$\underline{0164736}$    &0.39863$\underline{1384564}$\\
$S(-2)$                 &16.8$\underline{408}$                &16.8$\underline{447}$             &                                &1.075$\underline{95}$           &1.076$\underline{28}$\\
$S(-4)$                 &97$\underline{4.832}$                &97$\underline{5.135}$             &                                &3.69$\underline{533}$           &3.69$\underline{658}$\\
$S(-6)$                 &6.4$\underline{0829}[4]$             &6.4$\underline{1083}[4]$          &                                &14.9$\underline{422}$           &14.9$\underline{479}$\\
$S(-8)$                 &4.4$\underline{8404}[6]$             &4.4$\underline{8620}[6]$          &                                &64.3$\underline{838}$           &64.4$\underline{113}$\\
$S(-10)$                &3.2$\underline{6108}[8]$             &3.2$\underline{6295}[8]$          &                                &288.$\underline{547}$           &288.$\underline{684}$\\
$S(-12)$                &2.4$\underline{3429}[10]$            &2.4$\underline{3592}[10]$         &                                &132$\underline{8.18}$           &132$\underline{8.87}$\\
$S(-14)$                &1.8$\underline{5141}[12]$            &1.8$\underline{5282}[12]$         &                                &62$\underline{32.21}$           &62$\underline{35.79}$\\
$S(-16)$                &1.4$\underline{2796}[14]$            &1.4$\underline{2918}[14]$         &                                &296$\underline{67.5}$           &296$\underline{85.9}$\\
$\eta_1$                &77.$\underline{1282}$                &77.$\underline{1354}$             &                                &4.$\underline{76034}$           &4.$\underline{76058}$\\
$S^T(-2)$               &$-2.7\underline{3075}$               &$-2.7\underline{3118}$            &                                &$-0.15\underline{6423}$         &$-0.15\underline{6454}$\\
$S^T(-4)$               &$-15\underline{4.670}$               &$-15\underline{4.702}$            &                                &$-0.54\underline{4987}$         &$-0.54\underline{5110}$\\
$S^T(-6)$               &$-9.8\underline{8199}[3]$            &$-9.8\underline{8458}[3]$         &                                &$-2.10\underline{990}$          &$-2.11\underline{044}$\\
$S^T(-8)$               &$-6.\underline{71603}[5]$            &$-6.\underline{71818}[5]$         &                                &$-8.\underline{71572}$          &$-8.\underline{71824}$\\
$S^T(-10)$              &$-4.\underline{73561}[7]$            &$-4.\underline{73742}[7]$         &                                &$-37.\underline{5083}$          &$-37.\underline{5206}$\\
$S^T(-12)$              &$-3.\underline{42263}[9]$            &$-3.\underline{42417}[9]$         &                                &$-16\underline{6.164}$          &$-16\underline{6.226}$\\
$S^T(-14)$              &$-2.\underline{51902}[11]$           &$-2.\underline{52034}[11]$        &                                &$-7\underline{52.681}$          &$-7\underline{53.003}$\\
$S^T(-16)$              &$-1.\underline{88083}[13]$           &$-1.\underline{88197}[13]$        &                                &$-34\underline{71.37}$          &$-34\underline{73.06}$\\
$\eta^T_1$              &74.$\underline{6651}$                &74.$\underline{6713}$             &                                &4.$\underline{61201}$           &4.$\underline{61228}$\\
\end{tabular}
\end{ruledtabular}
\end{table*}
\endgroup

The relative error in the analytic representation of
$\alpha_1(\omega)$ for the $2\,^2\!P$ state of the Li atom is less
than 0.001\% for $\omega \leqslant 0.082$ a.u., 0.01\% for $\omega
\leqslant 0.0871$ a.u., and 0.1\% for $\omega \leqslant 0.0906$
a.u.. The relative error of the analytic representation for
$\alpha_1^T(\omega)$ is less than 0.001\% for $\omega \leqslant
0.0815$ a.u., 0.01\% for $\omega \leqslant 0.0932$ a.u., and 0.1\%
for $\omega \leqslant 0.0995$ a.u..

The dynamic polarizability of the Be$^+$ $2\,^2\!P$ state maintains
its accuracy over a larger range of $\omega$. It is accurate to
0.001 a.u. for $\omega\leqslant 0.3513$ a.u., to 0.01 a.u. for
$\omega \leqslant 0.3837$ a.u. and 0.1 a.u. for $\omega \leqslant
0.4111$ a.u.. The absolute error for $\alpha_1^T(\omega)$ is 0.001
a.u. for $\omega\leqslant 0.3788$ a.u., 0.01 a.u. for
$\omega\leqslant 0.4077$ a.u., and 0.1 a.u. for $\omega\leqslant
0.4291$ a.u..

The relative error between the analytic representation and Hylleraas
values of $\alpha_1(\omega)$ for the Be$^+$ $2\,^2\!P$ is less than
0.001$\%$ for $\omega \leqslant 0.332$ a.u., 0.01\% for $\omega
\leqslant 0.3536$ a.u. and $0.1\%$ for $\omega \leqslant 0.3708$
a.u.. The relative error for $\alpha_1^T(\omega)$ of the Be$^+$
$2\,^2\!P$ state is less than 0.001\% for $\omega \leqslant 0.3265$
a.u., 0.01\% for $\omega \leqslant 0.3434$ a.u. and 0.1\% for
$\omega \leqslant 0.3534$ a.u..

\section{Finite mass corrections}

The effect of the finite mass was to decrease the Li atom and Be$^+$
ion binding energies listed in the Table~\ref{tab:1}. Therefore, it is
not surprisingly that the $\omega = 0$ polarizabilities of the Li and
Be$^+$ ground states are increased in Tables ~\ref{tab:2} and
\ref{tab:3}. The overall changes of the $\omega = 0$ polarizabilities
are $0.03\%$ and $0.04\%$ for Li and Be$^+$ respectively. The finite
mass polarizabilities are larger than
the infinite mass values at $\omega = 0$ a.u.. These differences can
be taken as indicative of the overall change in the polarizabilities
at finite frequencies below the first excitation threshold. The
differences are naturally larger near thresholds.

The finite mass effect for the Li $2\,^2\!P$ state increased its
polarizability by 0.001$\%$ (Table \ref{tab:4}) while decreasing the
polarizability for the Be$^+$ $2\,^2\!P$ state (Table \ref{tab:4})
by 0.08$\%$. This behavior for Be$^+$ is due to the $2\,^2\!P$ $\to$
$2\,^2\!S$ downward transition. The increased negative contribution
from this transition is enough to outweigh the increased positive
contributions from transitions to more highly excited states.

As a general rule, the magnitude of the polarizabilities for both
upward and downward transitions increase for the finite mass
calculations. The residual Cauchy moments $S(-n)$ in
Table~\ref{tab:6} and Table~\ref{tab:7} all increase for the finite
mass calculations since these are computed exclusively from upward
transitions.

\section{Other effects and Uncertainties}

\subsection{Estimate of Relativistic effects}

The major omission from the present calculation is the inclusion of
relativistic effects. The larger part of the energy difference
between the present finite mass calculations and the experimental
binding energies in Table \ref{tab:1} is due to the omission of
relativistic effects. Relativistic effects will alter the
polarizability calculation in two ways. First, the energy
differences will be changed. Generally, the binding energies of all
states can be expected to be slightly larger. Secondly, there will
be some changes in the reduced matrix elements. The wave functions
for the $n\,^2\!S$ and $n\,^2\!P$ states can be expected to be
slightly more compact since they are more tightly bound.

Correcting for the relativistic energy is simply a matter of
replacing the theoretical energies in the sum rules by the
experimental values. The spin-orbit weighted averages were used for
states with $L\geq 1$. The corrections to the transition matrix
elements are made by recourse to calculations using a semi-empirical
model potential that supplements the potential field of a frozen
Hartree-Fock (HF) core with a tunable polarization potential
\cite{zhang,tang2,mitroy}. Polarizabilities for Li and Be$^+$
computed with this approach reproduce Hylleraas calculation at the
0.1$\%$ accuracy level~\cite{zhang,tang,tang2}.

The method used to estimate the relativistic effect upon matrix elements
relies on comparing two very similar calculations. One calculation has
its polarization potentials tuned to reproduce the finite mass energies
of Table~\ref{tab:1}. The other calculation is tuned to give the
experimental energies. The matrix elements for the low lying
transitions that dominate the dynamic polarizabilities are then
compared. The differences between the ``finite-mass'' calculation
and the ``experimental'' calculations are then determined. These
changes in the matrix elements are then applied as corrections to
the set of Hylleraas matrix elements. The only matrix elements that
are changed are those involving transitions inside the $2\,^2\!L$
and $3\,^2\!L$ level space. Transitions to these states dominate the
$2\,^2\!S$ and $2\,^2\!P$ polarizabilities.  The actual change in
the Li $2\,^2\!S$ $\to$ $2\,^2\!P$ matrix element was a reduction
of 0.0054$\%$.  The reduction in the Be$^+$ $2\,^2\!S$ $\to$ $2\,^2\!P$
matrix element was 0.011$\%$.

Using the new set of corrected matrix elements gives a ground state
polarizability of 164.114 a.u. (Table \ref{tab:1}). This represents
a reduction of the polarizability by 0.047 a.u..  A coupled cluster
calculation of the lithium ground state estimated that relativistic
effects reduced its polarizability by 0.06 a.u. \cite{lim}. The
static polarizability of the $2\,^2\!P$ state of $^7$Li, namely
126.947 a.u., was increased to 126.970 a.u. (Table \ref{tab:4}).
This gives a Stark shift of $-$37.144 a.u., which is in agreement
with the experiment of Hunter {\em et al} \cite{hunter} which gave
$-$37.14(2) for the $^7$Li $2\,^2\!S -2\,^2\!P_{1/2}$ Stark shift.
Another calculation was made to check the
$2\,^2\!P_{1/2}$:$2\,^2\!P_{3/2}$ polarizability difference. The
MBPT-SD calculation gave a difference of 0.015 a.u. \cite{johnson}.
Doing two calculations tuned to give a $2\,^2\!P$ spin-orbit
splitting of $1.77 \times 10^{-6}$ a.u.. (the energy splitting in
the MBPT-SD calculation \cite{johnson}) gave a polarizability
difference of 0.0145 a.u.. A further test was made by examination of
the line strengths of the Si$^{3+}$ $3\,^2\!S - 3\,^2\!P$ spin-orbit
doublet. A MBPT-SD calculation gave a line strength ratio of
1.000524 (once angular momentum factors were removed)
\cite{mitroy09a}. Turning the core potential in a semi-empirical
model based on the Schrodinger equation \cite{mitroy09a} gave a
value of 1.000618. The available evidence supports the conjecture
that it is possible to use the energy differences between the
Hylleraas and experimental energies to get an initial estimate of
relativistic corrections for other properties such as the
polarizability. The uncertainty in the correction would seem to be
about 20$\%$. To a certain extent the cancelations involved in
adding the finite mass and relativistic corrections together leads
to polarizabilities that are close to the infinite mass
polarizabilities.

The static polarizability of the Be$^+$ ground state was reduced
from 24.506 to 24.489 a.u.. This represents a reduction of 0.4$\%$.
However, the static scalar polarizability of the $2\,^2\!P$ state
increased from 2.0231 to 2.0285 a.u., an increase of 0.24$\%$. The
static tensor polarizability changed from 5.8589 to 5.8528 a.u.. The
heavier mass and larger nuclear charge means relativistic effects
are substantially larger than finite mass corrections.

Dynamic polarizabilities and their analytic representations from the
set of matrix elements with the estimate of the relativistic effect
are listed in Tables \ref{tab:2} - \ref{tab:7} as the recommended
values. The changes to analytic representation only involved changes
in the oscillator strength and energy differences for a few states.

\begin{table}[tbh]
\caption{Experimental $C_3$ values from analysis of the Li$_2$ spectrum and
$C_3$ values from the Hylleraas calculations. }
\label{tab:8}
\begin{ruledtabular}
\begin{tabular}{lc}
Source &  Value \\
\hline
Li$_2$ spectrum \cite{mcalexander}, $C_6$ fixed from \cite{yan1} &  11.0022(24)  \\
Li$_2$ spectrum \cite{leroy}, $C_6$ fixed from \cite{yan1} &  11.00241(23)  \\
Li$_2$ spectrum \cite{leroy}, $C_6$ fixed from \cite{zhang} &  11.00240(23)\\
Hylleraas $^{\infty}$Li & 11.000221 \\
Hylleraas $^{7}$Li & 11.001853 \\
Hylleraas $^{7}$Li: Recommended & 11.0007 \\
\end{tabular}
\end{ruledtabular}
\end{table}

\subsection{The $2\,^2\!S$ $\to$ $2\,^2\!P$ matrix element and uncertainties}

Recently Le~Roy {\em et al} \cite{leroy} analysed the ro-vibrational
spectrum of the lithium dimer obtaining an estimate for the $C_3$
parameter describing the long range $C_3/R^3$ potential of the
A-state that dissociates to the $2\,^2\!S$ and $2\,^2\!P$ states.
The $C_3$ parameter can be related to the $2\,^2\!S - 2\,^2\!P$
multiplet strength. The determination of Le~Roy {\em et al}
represented an order of magnitude improvement in precision over any
previous determination of $C_3$.

The current value of $C_3 = 11.0007$ a.u. computed with relativistic
corrections is about 0.0155$\%$ smaller than the experimental value
of Le~Roy {\em et al}. The finite mass calculation with the
relativistic correction is closer to experiment than the infinite
mass $C_3$, but there is a remaining discrepancy of 0.0017 a.u.. It
is not likely that QED effects can explain the discrepancy as
Pachucki {\em et al} found that these were 2.5 times smaller than
relativistic effects in the polarizability of helium
\cite{pachucki}. It must be recalled that the Le~Roy {\em et al}
experiment is reporting an order of magnitude improvement in
experimental precision. Going to such extreme levels of precision
means there might be small corrections that need to be applied to
the analysis of the data that have not received consideration. For
example, the value of $C_3$ will be different for states asymptotic
to the $2\,^2\!P_{1/2}$ and $2\,^2\!P_{3/2}$ levels. The analysis of
Le~Roy {\em et al} uses a common $C_3$ value for both members of the
spin-orbit doublet. Irrespective of this, it should be noted that
experiment and theory are incompatible at precisions better than
0.01$\%$.

The difference between the present and Le~Roy $C_3$ is used to
assign an error to the present polarizability calculation. Changes
of 0.008$\%$ were made to the $2(3)\,^2\!S - 2(3)\,^2\!P$ matrix
elements, the polarizabilities were recomputed, and the differences
assigned as the uncertainty in the recommended values. This
difference is actually larger than the estimated relativistic change
in the matrix element. Therefore, the recommended static
polarizability of the Li ground state is 164.11(3) a.u..
Uncertainties in the $2(3)\,^2\!P - 3\,^2\!D$ matrix elements are
smaller (relativistic effects have a smaller impact on these matrix
elements) and have not been included in the uncertainty analysis.
The final value for the static scalar polarizability of the
$2\,^2\!P$ state was 126.970(4) a.u., while the tensor
polarizability was 1.612(4) a.u..

The same uncertainty analysis was applied to the Be$^+$ ion
polarizabilities. The recommended value for the ground state is
24.489(4) a.u.. The static $2\,^2\!P$ scalar polarizability was set
as 2.0285(10) a.u. while the tensor polarizability was set to
5.8528(10) a.u..

Uncertainties in the recommended dynamic polarizabilities in Tables
\ref{tab:2}, \ref{tab:3}, \ref{tab:4} and \ref{tab:5} were computed
by making corrections to the matrix elements, recomputing and then
observing the change. These uncertainties are best interpreted as
indicative as opposed to rigorous estimates.

\section{Summary}

Definitive non-relativistic values for the dynamic dipole
polarizabilities of Li and Be$^+$ in their low-lying
$2\,^2\!S$ and $2\,^2\!P$ states have been established using the
variational method with Hylleraas basis sets. Calculation for both
finite and infinite nuclear mass systems have been
performed. Analytic representations for the dynamic polarizabilities
of the Li atom and the Be$^+$ ion have also been developed. These results
can serve as a standard against which any other calculation can be
judged.

Subsidiary calculations have been used to estimate the impact of
relativistic effects that are not explicitly included in the
Hylleraas calculation. It is recommended that the value of 164.11(3)
a.u. be adopted as the static polarizability of $^7$Li. The
uncertainty of 0.03 a.u. is based on the difference between the
present $C_3$ and that of Le Roy {\em et al} \cite{leroy}. This
accuracy level is also supported by the $^7$Li
$2\,^2\!S-2\,^2\!P_{1/2}$ Stark shift of $-$37.14 a.u. which is in
perfect agreement with the value of $-$37.14(2) given by the high
precision experiment of Hunter {\em et al} \cite{hunter}. The
recommended static polarizability for Be$^+$ is 24.489(4) a.u..

The dynamic polarizabilities that have been obtained can be used as
an atom based standard for electromagnetic field intensity. These
polarizabilities can be regarded as an initial attempt to develop
atom based standards for polarizability and Stark shift
measurements. The primary virtue of the method with which the
relativistic corrections were evaluated was simplicity of
computation. For present purposes, the estimate of the relativistic
corrections only has to be accurate to 10-20$\%$ for the recommended
polarizabilities to be valid. The comparisons that have been done
with fully relativistic calculations suggest that the estimates of
the relativistic corrections are indeed accurate at this level.
However, a more rigorous estimate using the Briet-Pauli Hamiltonian
and perturbation theory would be desirable \cite{pachucki,cencek,lach}.

\begin{acknowledgments}
This work was supported by NNSF of China under Grant No. 10974224
and by the National Basic Research Program of China under Grant No.
2010CB832803. Z.-C.Y. was supported by NSERC of Canada and by the
computing facilities of ACEnet, SHARCnet, WestGrid, and in Part by
the CAS/SAFEA International Partnership Program for Creative
Research Teams. J.M. would like to thank the Wuhan Institute of
Physics and Mathematics for its hospitality during his visit.  We
would like thank M.S.Safronova and J.F.Babb for useful communications
during this work.
\end{acknowledgments}


\end{document}